\pdfoutput=1
\documentclass[aps,prb,twocolumn,superscriptaddress,a4paper,
    floatfix]{revtex4}
\usepackage[utf8]{inputenc}
\usepackage{graphicx}
\usepackage{amsmath,amssymb}
\usepackage{hyperref}

\def\92{$|{-9/2}\rangle$}
\def\72{$|{-7/2}\rangle$}
\def\52{$|{-5/2}\rangle$}

\begin{document}
\title{A Mott insulator of fermionic atoms in an optical lattice}

\author{Robert J\"ordens}
\altaffiliation{These authors contributed equally to this work.}
\affiliation{Institute for Quantum Electronics, ETH Zurich,
8093 Zurich, Switzerland}
\author{Niels Strohmaier}
\altaffiliation{These authors contributed equally to this work.}
\affiliation{Institute for Quantum Electronics, ETH Zurich,
8093 Zurich, Switzerland}
\author{Kenneth G\"unter}
\affiliation{Institute for Quantum Electronics, ETH Zurich,
8093 Zurich, Switzerland}
\affiliation{Laboratoire Kastler Brossel, \'Ecole Normale Sup\'erieure,
24 Rue Lhomond, 75005 Paris, France}
\author{Henning Moritz}
\affiliation{Institute for Quantum Electronics, ETH Zurich,
8093 Zurich, Switzerland}
\author{Tilman Esslinger}
\affiliation{Institute for Quantum Electronics, ETH Zurich,
8093 Zurich, Switzerland}

\date{September 11, 2008}

\maketitle

{\bf

  In a solid material strong interactions between the electrons can lead
  to surprising properties. A prime example is the Mott insulator, where
  the suppression of conductivity is a result of interactions and not
  the consequence of a filled Bloch band\cite{Mott}. The proximity to
  the Mott insulating phase in fermionic systems is the origin for many
  intriguing phenomena in condensed matter physics\cite{Imada1998}, most
  notably high-temperature superconductivity\cite{Lee2006}. Therefore it
  is highly desirable to use the novel experimental tools developed in
  atomic physics to access this regime.  Indeed, the Hubbard
  model\cite{Hubbard}, which encompasses the essential physics of the
  Mott insulator, also applies to quantum gases trapped in an optical
  lattice\cite{Jaksch1998, Hofstetter2002}. However, the Mott insulating
  regime has so far been reached only with a gas of
  bosons\cite{Greiner2002}, lacking the rich and peculiar nature of
  fermions. Here we report on the formation of a Mott insulator of a
  repulsively interacting two-component Fermi gas in an optical lattice.
  It is signalled by three features: a drastic suppression of doubly
  occupied lattice sites, a strong reduction of the compressibility
  inferred from the response of double occupancy to atom number
  increase, and the appearance of a gapped mode in the excitation
  spectrum. Direct control of the interaction strength allows us to
  compare the Mott insulating and the non-interacting regime without
  changing tunnel-coupling or confinement. Our results pave the way for
  further studies of the Mott insulator, including spin ordering and
  ultimately the question of $d$-wave superfluidity\cite{Hofstetter2002,
  Trebst2006}.

}

The physics of a Mott insulator is well captured by the celebrated
Hubbard model which is widely used to describe strongly interacting
electrons in a solid. It assumes a single static energy band for the
electrons and local interactions, i.\,e. spin-up and spin-down fermions
are moving on a lattice and interact when occupying the same lattice
site. The consequence of strong repulsive interactions is that even
fermions in different spin states tend to avoid each other.  In the case
of a half filled band the particles get localised and an incompressible
state with one fermion per site forms. Since no symmetry is broken, the
transition between the metallic and the Mott insulating regime at finite
temperature exhibits a crossover rather than a phase transition.

The Hubbard model ignores various complexities of
materials\cite{Imada1998} but it has been highly successful to study the
nature of the Mott insulating regime, including magnetic
phenomena\cite{Imada1998} and high-temperature
superconductivity\cite{Lee2006}. Yet, despite its simplicity, it turned
out that the fermionic Hubbard model is in many cases computationally
intractable and important puzzles remain to be solved.  In particular,
the question whether the ground state of the lightly doped 2D Hubbard
model supports $d$-wave superconductivity is as yet unanswered.

Compared to real materials, a fermionic quantum gas trapped in an
optical lattice is a much purer realisation of the Hubbard
model\cite{Jaksch1998, Hofstetter2002, Bloch2007, Stringari2008,
Georges2006}.  It offers a new approach to understand the physics of
strongly correlated systems. In an optical lattice three mutually
perpendicular standing laser waves create a periodic potential for the
atoms.  The kinetics of the atoms is determined by their tunnelling rate
between neighbouring lattice sites and the interaction is due to
inter-atomic collisions occurring when two atoms are on the same site.
In a gas of fermions in different spin states this collisional
interaction can be widely tuned through a Feshbach resonance without
encountering significant atom losses\cite{Stringari2008}.

A landmark result has been the observation of the superfluid to Mott
insulator transition using bosonic atoms trapped in an optical
lattice\cite{Greiner2002}. Yet it is the fermionic character combined
with repulsive interactions which provides the intimate link to
fundamental questions in strongly correlated electron systems. Whereas
experimental studies of fermionic quantum gases in three-dimensional
optical lattices have been scarce and focused on non-interacting and
attractively interacting cases\cite{Koehl2005, Stoeferle2006, Chin2006,
Rom2006, Strohmaier2007}, we investigate the repulsive Fermi-Hubbard
model and its paradigm, the Mott insulator.

\begin{figure}
  \includegraphics[width=1\columnwidth]{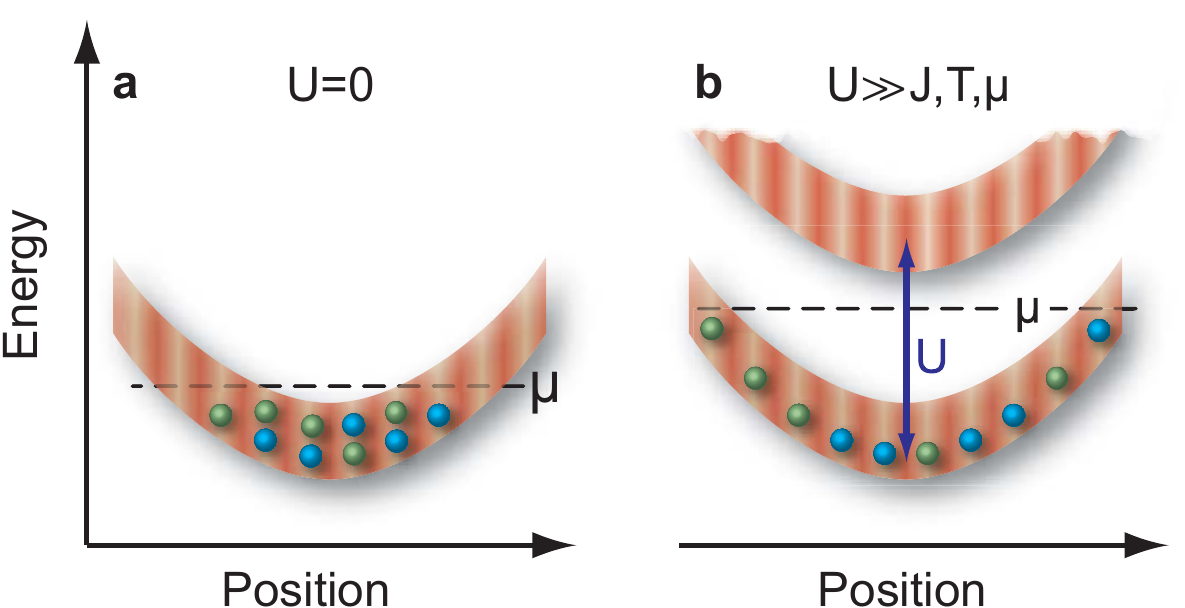}

  \caption{{\bf Energy spectrum of a Fermi gas in an optical lattice
  with an underlying confining potential.} In the non-interacting case
  (\textbf{a}) the curvature of the lowest Bloch band reflects the
  harmonic confinement.  At zero temperature all states up to the
  chemical potential $\mu$ are filled with atoms of both spin states
  (green and blue). \textbf{b}, In the Mott insulating limit the energy
  cost for creating doubly occupied sites greatly exceeds the
  temperature $T$ and the kinetic energy parametrised by $J$, giving
  rise to a gap of order $U$. The energy spectrum of single particle
  excitations is then depicted by two Hubbard bands. Doubly occupied
  sites correspond to atoms in the upper Hubbard band.}
  
  \label{fig:energy-scheme}
\end{figure}

In optical lattice experiments the presence of an underlying harmonic
trapping potential has an important influence on the observable physics.
Let us first consider a zero temperature Fermi gas prepared in an equal
mixture of two non-interacting spin components. All available single
particle quantum states will be filled up to the Fermi energy and, for a
sufficiently large number of trapped atoms, a band insulating region
with two atoms per site appears in the trap centre, surrounded by a
metallic shell with decreasing filling, see figure~1. An important
quantity to characterise the state of the system is the fraction $D$ of
atoms residing on lattice sites which are occupied by two atoms, one
from each component. For the non-interacting case this double occupancy
should increase in a continuous fashion with the number $N$ of atoms in
the trap.

A very different scenario can be anticipated for a gas with increasingly
strong repulsive interactions. A Mott insulator will
appear\cite{Rigol2003, Helmes2008}, at first in those regions of the
trap where the local filling is approximately one atom per site. For
very strong repulsion the entire centre of the trap contains a Mott
insulating phase and double occupancy is suppressed, see figure~1. Since
the Mott insulating region is incompressible\cite{Georges1996,
Rigol2003}, the suppression of double occupancy should be robust against
a tightening of the trapping potential, or equivalently, against an
increase in the number of trapped atoms. However, once the chemical
potential $\mu$ reaches a level where double occupation of sites becomes
favourable, a metallic phase appears in the centre and the double
occupancy increases accordingly. The energy spectrum in the Mott
insulating phase is gapped, with a finite energy cost required to bring
two atoms onto the same lattice site. This energy has to be large
compared to the temperature in order to keep the number of thermally
excited doubly occupied sites small. Thermally excited holes in the
centre are suppressed by the chemical potential $\mu$.

Our experiment is performed with a quantum degenerate gas of fermionic
$^{40}$K atoms, prepared in a balanced mixture of two magnetic sublevels
of the $F=9/2$ hyperfine manifold ($F$: total angular momentum).
Feshbach resonances allow us to tune the $s$-wave scattering length
between $a=240\pm4a_0$ and $810\pm40a_0$ as well as to prepare
non-interacting samples. Here $a_0$ is the Bohr radius.  The
two-component Fermi gas is subjected to the potential of a
three-dimensional optical lattice of simple cubic symmetry. In terms of
the recoil energy $E_R=h^2/(2m\lambda^2)$ the lattice potential depth
$V_0$ is chosen between $6.5$ and $12E_R$. Here $h$ is Planck's
constant, $m$ the atomic mass and $\lambda=1064\,\text{nm}$ the
wavelength of the lattice beams. The system is described by the Hubbard
Hamiltonian:

$$ \hat H= -J\sum_{\langle ij\rangle,\sigma}
    (\hat c^\dagger_{j\sigma}\hat c_{i\sigma}+\text{h.c.}) +
    U\sum_i \hat n_{i\uparrow}\hat n_{i\downarrow} +
    \sum_i\epsilon_i\hat n_i.
\label{eq:fermi-hubbard} $$

The onsite interaction energy is given by $U$ and the tunnelling matrix
element between nearest neighbours $\langle ij\rangle$ by $J$. The
quotient $U/(6J)$ which characterises the ratio between interaction and
kinetic energy can be tuned from zero to a maximum value of 30. The
fermionic creation operator for an atom on the lattice site $i$ is given
by $\hat c^\dagger_{i\sigma}$, where $\sigma\in\{\uparrow,\downarrow\}$
denotes the magnetic sublevel.  The particle number operator is $\hat
n_i=\hat n_{i\uparrow}+\hat n_{i\downarrow}$, $\hat n_{i\sigma}=\hat
c^\dagger_{i\sigma}\hat c_{i\sigma}$, and $\epsilon_i$ is the energy
offset experienced by an atom on lattice site $i$ due to the harmonic
confining potential.

In order to characterise the state of the Fermi gas in the optical
lattice we have developed a technique to measure the fraction $D$ of
atoms residing on doubly occupied sites with a precision of down to 1\%.
The experimental procedure is as follows. The depth of the optical
lattice is rapidly increased to $30E_R$ to inhibit further tunnelling.
In the next step, we shift the energy of atoms on doubly occupied sites
by approaching a Feshbach resonance. This enables us to specifically
address only atoms on doubly occupied sites by using a radio frequency
pulse to transfer one of the spin components to a third, previously
unpopulated magnetic sublevel. The fraction of transferred atoms is
obtained from absorption images and allows us to deduce the double
occupancy. 

\begin{figure}
  \includegraphics[width=1\columnwidth]{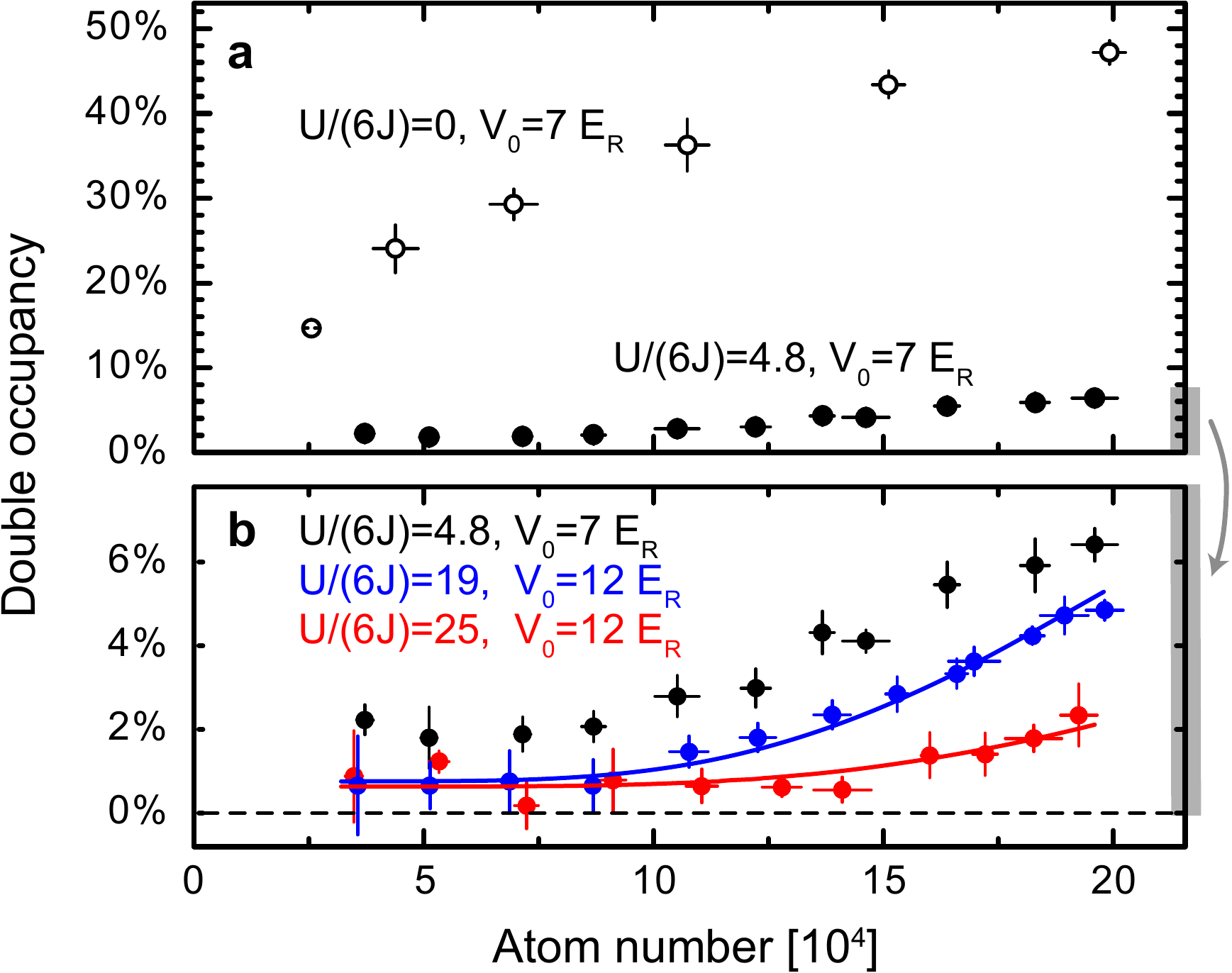}

  \caption{\textbf{Double occupancy in the non-interacting and Mott
  insulating regime.} \textbf{a,} A significant increase of the double
  occupancy with atom number is observed in the non-interacting regime
  (empty circles) whereas on entering the Mott insulating regime the
  double occupancy is suppressed (filled circles). The corresponding
  onsite interaction strengths are $U/h=0\pm80\,\text{Hz}$ and
  $U/h=5.0\pm0.6\,\text{kHz}$.  \textbf{b,} In the Mott insulating
  regime the double occupancy is strongly suppressed.  It starts to
  increase for large atom numbers indicating the formation of a metallic
  region in the trap centre. The blue and red lines represent the
  theoretical expectation for $D$ in the atomic limit (see text and
  methods summary).  Values and error bars are the mean and s. d. of 4
  to 8 identical measurements. The systematic relative errors for the
  atom number, double occupancy, and lattice depth are estimated to be
  20\%, 10\%, and 10\% respectively, with corresponding relative errors
  in $J$ of up to 30\%. These systematic errors apply to all further
  measurements.}
  
  \label{fig:double-occupancy}
\end{figure}

The double occupancy as a function of total atom number is plotted in
figure~2a, where the non-interacting situation is compared to the case
of strong repulsive interactions. The former shows the expected rapid
increase of double occupancy with atom number.  A strikingly different
behaviour is observed in the strongly repulsive regime with $U\gg
J,T,\mu$, where a Mott insulator is expected.  The double occupancy is
strongly reduced to values systematically below 2\% for small atom
numbers. This is direct evidence for the suppression of fluctuations in
the occupation number and for the localisation of the atoms.

\begin{figure}
  \includegraphics[width=1\columnwidth]{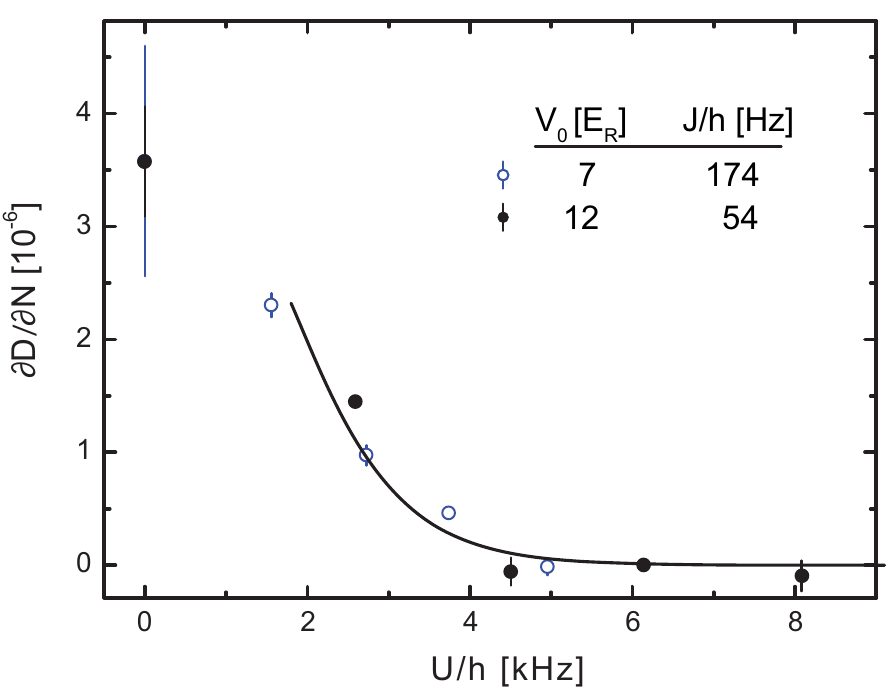}

  \caption{{\bf The transition to an incompressible sample.} Upon
  changing $U$, two regimes can be distinguished by the slope $\partial
  D/\partial N$. For vanishing interaction the large initial slope
  signals the filling of the Bloch band. Increasing $U$ reduces the
  double occupancy. For $U/h\gtrsim 5\,\text{kHz}$ a change in atom
  number can no longer change the double occupancy.  The compressibility
  $\partial D/\partial N$ is obtained from a least squares fit of
  $D(N)=(\partial D/\partial N)N+D_0$ to data such as shown in figure~2,
  with atom numbers in the interval from $25\times10^3$ to
  $8\times10^4$.  Error bars denote the confidence interval of the fit.
  The expected slope in the atomic limit is illustrated with a black
  line for a lattice depth of $12E_R$ and $T=0.28T_F$.
  }
  
  \label{fig:compressibility}
\end{figure}

In order to experimentally investigate the compressibility on entering
the Mott insulating regime we determine how the double occupancy changes
with increasing atom number, i.e.~we extract the slope $\partial
D/\partial N$ from curves such as shown in figure~2. This slope is a
good measure of the compressibility $\kappa=\partial n/\partial \mu$ in
those regions of the cloud where the filling $n$ is near unity or
larger, since $n$ increases with $D$. We estimate the filling in the
trap centre for the non-interacting case from the measured double
occupancy\cite{Koehl2006}. It significantly exceeds one atom per site,
e.\,g. $\langle \hat{n} \rangle=1.4$ for $N=5\times 10^4$, $V_0=7E_R$
and a temperature $T$ of 30\% of the Fermi temperature $T_F$.

The slope $\partial D/\partial N$ is displayed in figure~3 for a wide
range of interaction strengths. The data shows that we access two
regimes: For small onsite interaction energies $U$ the slope $\partial
D/\partial N$ is positive and the system is compressible. Yet for
$U/h>5\,\text{kHz}$ the measured compressibility vanishes. This
indicates that we have entered the Mott insulating regime. It implies a
large central region with a filling reduced to one atom per site,
surrounded by a metallic region with lower filling.

Further insight is gained by comparing our measurements with the
theoretical values of $\partial D/\partial N$ calculated in the atomic
limit\cite{Gebhard1997} of the Hubbard model, including confinement and
finite temperature. In this limit the kinetic energy is neglected by
setting the tunnelling matrix element $J$ to zero (see also methods
summary). We find good agreement between theory (black line in figure~3)
and experimental data for $U\gg 6J$, where the above assumption is
acceptable. For the calculation we have assumed a temperature of
$T=0.28T_F$, which is deduced from the entropy in the dipole trap as
discussed in the methods section. For zero temperature the slope
$\partial D/\partial N$ would vanish as soon as $U$ becomes larger than
the chemical potential $\mu$, which is $h\times2.7\,\text{kHz}$ for
$N=8\times10^4$ atoms and a lattice potential of $V_0=12E_R$. Both our
measurements and the model at finite temperature show a finite
compressibility extending beyond $U/h=2.7\,\text{kHz}$, which can be
attributed to thermal excitations. For the largest attained interaction
$U/h=8.1\,\text{kHz}$ the thermal excitations are characterised by
$T=0.11U/k=0.28T_F$ corresponding to 3\% vacancies in the trap centre
according to the theoretical analysis presented in the methods section
($k$ is Boltzmann's constant). The vanishing slope $\partial D/\partial
N$ at this filling implies incompressibility of the core. The obtained
ratio $T/U$ is comparable to estimates for the bosonic Mott
insulator\cite{Gerbier2007}.

In the strongly repulsive regime, the measured compressibility should
vanish if $\mu<U$. For atom numbers corresponding to higher chemical
potentials a metallic phase will appear in the trap centre and the
double occupancy will increase. We observe this characteristic
behaviour\cite{Gerbier2006} which is a consequence of the presence of a
Mott insulator, see figure~2b. The behaviour agrees well with that
expected from the Hubbard model in the atomic limit (lines in
figure~2b). The free parameters in the theory curves, the temperature
and a constant offset in $D$, are determined by a least squares fit to
the data. The fits yield temperatures of $0.2\pm0.1T_F$. However, the
accuracy is limited due to the high sensitivity to the energy gap and
the harmonic confinement. The constant offset in $D$ accounts for the
finite double occupancy in the ground state caused by second order
tunnelling processes as well as a systematic offset of 0.5\% stemming
from technical imperfections in the initial preparation of the spin
mixture.

\begin{figure}
  \includegraphics[width=1\columnwidth]{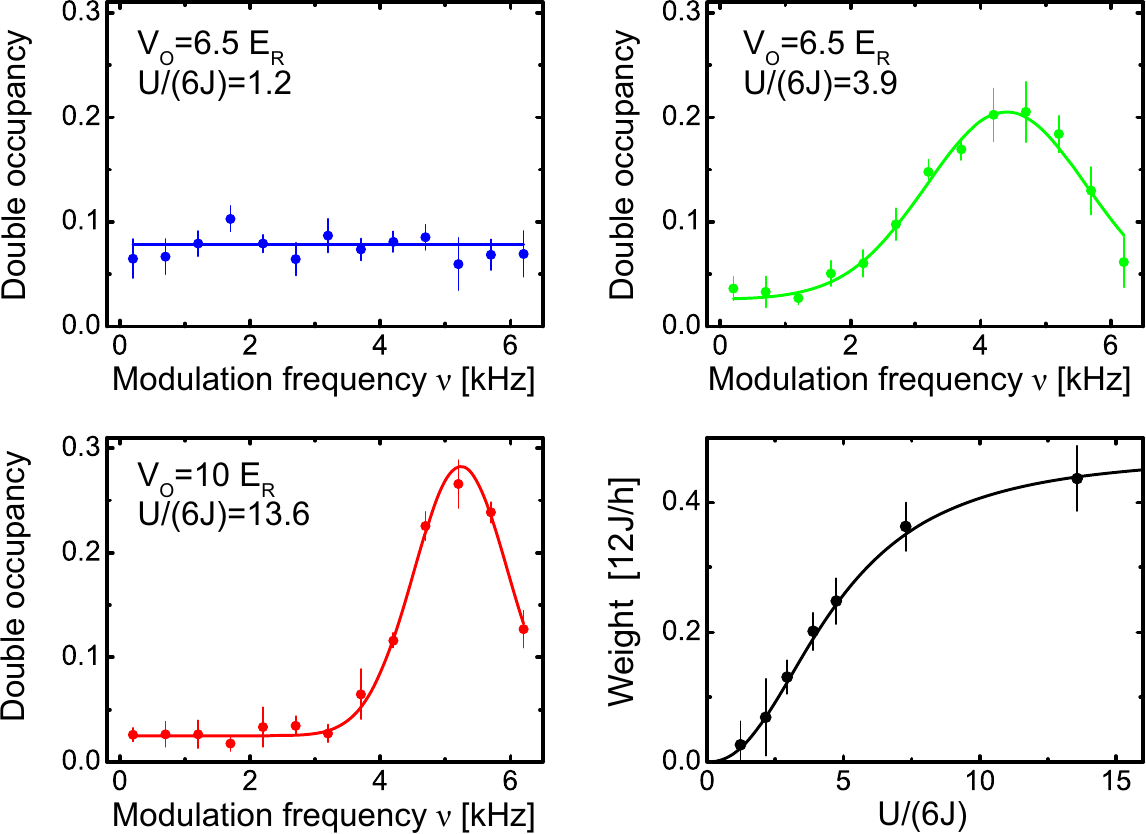}

  \caption{{\bf Emergence of a gapped mode.} With increasing interaction
  (blue, green, red) the response to modulation of the lattice depth
  shows the appearance of a gapped mode.  The weight of this peak grows
  with $U/(6J)$ and saturates.  All modulation spectra were obtained
  with $(32\pm 2)\times10^3$ atoms.  The weight of the peak shown in the
  bottom right panel is $\sum_i\Delta \nu \left[ D(\nu_i) -
  \frac{1}{2}(D(200\,\text{Hz}) + D(700\,\text{Hz}))\right]$, where
  $D(\nu_i)$ is the measured double occupancy at frequencies $\nu_i$
  which are evenly spaced in steps of $\Delta\nu=500\,\text{Hz}$. It is
  plotted in units of $12J/h$. The first four datapoints are taken for a
  lattice depth of $6.5E_R$, the next at 7, 8 and $10E_R$, from left to
  right, respectively. The lines serve as a guide to the eye. Values and
  error bars are the mean and s.~d.~of 4 to 8 identical measurements.}

  \label{fig:modulation}
\end{figure}

An important feature of a Mott insulator is the energy gap in the
excitation spectrum. The lowest lying excitations are particle-hole
excitations centred at an energy $U$. The actual gap in the energy
spectrum is reduced with respect to this value due to the width of the
energy bands experienced by particles and holes\cite{Brinkman1970}. A
suitable technique for probing this excitation spectrum is to measure
the response of the quantum gas in the optical lattice to a modulation
of the lattice depth\cite{Stoeferle2004, Kollath2007, Huber2008}: we
apply 50 cycles of sinusoidal intensity modulation of all three lattice
beams with an amplitude of 10\%. The response is quantified by recording
the double occupancy as a function of modulation frequency.  With
increasing interactions we observe the emergence of a gapped mode in the
excitation spectrum (figure~\ref{fig:modulation}). For small values of
$U/(6J)$, the double occupancy is not affected by the modulation of the
lattice depth but for large values of $U/(6J)$ a distinct peak appears
for modulation frequencies $\nu$ near $U/h$. Furthermore, the area under
the excitation curve divided by 12 $J/h$ as a measure for its width
increases with interaction strength and starts to saturate at large
values of $U/(6J)$, see figure~\ref{fig:modulation}d.

The presented approach to the physics of the repulsive Fermi-Hubbard
model is completely different and complementary to that encountered in
solid-state systems, and provides a new avenue to one of the predominant
concepts in condensed matter physics. In this first experiment we have
found clear evidence for the formation of a Mott insulator of fermionic
atoms in the optical lattice. We could set limits for the deviation from
unity filling in the Mott insulator by directly measuring the residual
double occupancy and by deducing the number of holes from a realistic
estimate of the temperature. The temperature is found to be small
compared to the onsite interaction energy and the Fermi temperature. In
addition, we have obtained good quantitative agreement with the Hubbard
model in the atomic limit for a wide range of parameters.  In further
investigations of e.g.~the energy spectra, the high resolution achieved
may give direct insights into the width of Hubbard
bands\cite{Brinkman1970}, the lifetime of excitations and the level of
anti-ferromagnetic ordering\cite{Georges2006,Kollath2007} in the system.

\subsection*{Methods summary}

In the atomic limit $U\gg 6J$ of the Hubbard model we assume full
localisation of the fermions and thus neglect the kinetic energy. Each
site is treated in the grand canonical ensemble with three possible
occupation numbers $n\in\{0,1,2\}$. The partition function for site $i$
is then $\mathcal{Z}_i=\sum_n z^n\exp(-\beta E_{i,n}) =1+2z\exp(-\beta
\epsilon_i)+z^2\exp(-2\beta \epsilon_i-\beta U)$ where $\beta=1/kT$ is
the inverse temperature, $z=\exp(\beta\mu)$ the fugacity, $\mu$ the
chemical potential, $E_{i,n}$ the energy of $n$ particles on site $i$,
and $\epsilon_i$ the energy offset due to the harmonic confinement. For
the probability to find a double occupancy $\langle d\rangle_i$ or a
vacancy $\langle v \rangle_i$ one obtains $\langle
d\rangle_i=z^2\exp(-2\beta \epsilon_i-\beta U)/\mathcal{Z}_i$ and
$\langle v \rangle_i=1/\mathcal{Z}_i$. Double occupancy $D$ and total
particle number $N$ of the system are obtained by summing over all
sites, e.\,g. $D=\sum_i 2\langle d\rangle_i/N$, where the equation for
$N$ is first solved numerically with respect to $z$. The entropy is
$S=\frac{\partial}{\partial T}(k T \sum_i\ln\mathcal{Z}_i)$.  We
calculate the temperature in the lattice by assuming that this entropy
is the same as the entropy determined from temperature measurements in
the dipole trap (see methods). The fits in figure~2b involve $U$ as
determined by modulation spectroscopy ($U/h=4.7\pm0.1\,\text{kHz}$ and
$6.1\pm0.1\,\text{kHz}$) since band structure calculations disagree with
the measured value by up to 30\% for the largest scattering lengths.

\bigskip

\paragraph*{Acknowledgements}

We thank J.~Blatter, S.~Huber, M.~K\"ohl, C.~Kollath, L.~Pollet,
N.~Prokof'ev, M.~Rigol, M.~Sigrist, M.~Troyer and W.~Zwerger for
discussions.  Funding was provided by the Swiss National Science
Foundation (SNF), the E.U.  projects Optical Lattices and Quantum
Information (OLAQUI) and Scalable Quantum Computing with Light and Atoms
(SCALA) and the Quantum Science and Technology (QSIT) project of ETH
Zurich.

\paragraph*{Author Information}

Correspondence should be addressed to H.~M.
(\url{mailto:moritz@phys.ethz.ch}).

\clearpage

\section*{Methods}

\paragraph*{Preparation.}

After sympathetic cooling with $^{87}$Rb, $2\times10^6$ fermionic
$^{40}$K atoms are transferred into a dipole trap operating at a
wavelength of $826\,\text{nm}$. Initially, a balanced spin mixture of
atoms in the $|m_F\rangle=$ \92 and \72 states is prepared and
evaporatively cooled at a magnetic bias field of 203.06~Gauss. Using
this mixture we realise non-interacting samples with a scattering length
of $a=0\pm10a_0$. Repulsive interactions are obtained by transferring
the atoms in the \72 state to the \52 state during the evaporation, thus
cooling and preparing a spin mixture of atoms in \92 and \52 states,
close to a Feshbach resonance at 224.21~Gauss\cite{Regal2003}. After
tuning the scattering length to the desired value we load the atoms into
the lowest Bloch band of the optical lattice by increasing the intensity
of three retroreflected laser beams within $200\,\text{ms}$ using a
spline ramp.  The beams have circular profiles with $1/e^2$ radii of
$(160, 180, 160)\,\mathrm{\mu m}$ at the position of the atoms. For a
given scattering length and lattice depth $J$ and $U$ are inferred from
the Wannier functions including the interaction induced coupling to the
second Bloch band. The latter leads to corrections of up to 15\% with
respect to the single band model.

\paragraph*{Radio-frequency spectroscopy.}

By increasing the depth of the optical lattice to $30E_R$ in 0.5\,ms
tunnelling is suppressed. The magnetic field is tuned to 201.28~Gauss,
where a molecular state for a \92, \72 pair with binding energy $h\times
99\pm1\,\text{kHz}$ and a weakly interacting state for a \92, \52 pair
exist\cite{Stoeferle2006}. A radio-frequency $\pi$-pulse dissociates
(associates) pairs and changes the spin state of those \72 (\52) atoms
that share a site with a \92 atom. Finally the magnetic field is
increased to 202.80~Gauss dissociating any molecules and the lattice
potential is ramped down in 10\,ms. All confining potentials are
switched off and the homogeneous magnetic bias field is replaced by a
magnetic gradient field in the same direction applied for 2\,ms, thus
spatially separating the spin states.

\paragraph*{Imaging.}

After 6\,ms of time-of-flight all three clouds are imaged
simultaneously.  Due to a reproducible change of the imaging beam
profile between the atomic absorption image and the subsequent reference
image without atoms, residual structures are present in the density
profiles. These are reduced by repeating the entire experiment without
loading atoms and subtracting the obtained residual density distribution
from the atomic density distribution. The number of atoms $N_{m_F}$ per
spin component $m_F$ is determined from the 2D column densities by
simultaneously fitting the sum of three quartic terms $A\cdot
\max(1-(x/w_x)^4,0)\cdot\max(1-(y/w_y)^4,0)$ with identical widths
$w_{x,y}$ and mutual distances. This permits accurate detection of atom
numbers down to 200 atoms per spin state. We have validated the absolute
accuracy of the fits against integration of the density. The fraction
$D$ of atoms residing on doubly occupied sites is defined as
$D=2N_{m_F\prime}/N$ where $N=N_{-9/2}+N_{-7/2}+N_{-5/2}$ and
$m_F\prime=-5/2$ ($-7/2$) for samples initially containing atoms in the
\72 (\52) states, respectively. The relative uncertainty in $D$ is 10\%,
validated against measurements of the adiabatic molecule formation
efficiency\cite{Stoeferle2006,Strohmaier2007}. We estimate the relative
systematic error for the total atom number $N$ to be less than 20\%. The
\92, \52 mixture shows an offset of 0.5\% in $D$ due to \72 atoms
remaining from the initial spin transfer during evaporation.

\paragraph*{Temperature.}

The temperature is measured in the harmonic dipole trap before ramping
up the lattice and after a subsequent reversed ramp back into the dipole
trap. The highest temperatures measured before and after ramping are
$T_i=0.15T_F$ and $T_f=0.24T_F$, respectively. Since we expect
non-adiabatic heating to occur during the lattice ramp up as well as
during ramp down, we use the mean value of $0.195T_F$ as a realistic
estimate. With this, we calculate a temperature of $T=0.28T_F$ in the
Mott insulating regime ($a=810a_0$, $V_0=12E_R$, $N=10^5$),
corresponding to 3.3\% holes and a compressibility as low as $\partial
n/\partial \mu=0.09/\mu$ in the centre. For the temperatures in the
dipole trap before and after the lattice ramp we would obtain 0.3\%
holes for $T_i$ and 11.5\% for $T_f$. The reported temperatures
represent upper limits, since we have achieved temperatures down to
$0.08T_F$ in the dipole trap prior to loading.  Due to inelastic
collisions we lose at most $(4.8\pm0.6) \%$ of the atoms during the
preparation of the Mott insulating state for the parameters above, where
the losses are expected to be highest. The inelastic decay time for
atoms on doubly occupied sites exceeds 850\,ms, which is significantly
longer than the relevant experimental timescale.


\begin{thebibliography}{99}

\bibitem{Mott} Mott, N.~F. \emph{Metal Insulator Transitions} (Taylor
  and Francis, London, 1990).

\bibitem{Imada1998} Imada, M., Fujimori, A. \& Tokura, Y.
  Metal-insulator transitions. \emph{Rev. Mod. Phys.} {\bf 70},
  1039-1263 (1998).

\bibitem{Lee2006} Lee, P.~A., Nagaosa, N. \& Wen, X.-G. Doping a Mott
  insulator: physics of high-temperature superconductivity. \emph{Rev.
  Mod. Phys.} {\bf 78}, 17-85 (2006).

\bibitem{Hubbard} Hubbard, J. Electron correlations in narrow energy
  bands. \emph{Proc. R. Soc. Lond. A} {\bf 276}, 238-257 (1963).

\bibitem{Jaksch1998} Jaksch, D., Bruder, C., Cirac, J.~I., Gardiner,
  C.~W. \& Zoller, P. Cold bosonic atoms in optical lattices.
  \emph{Phys.  Rev. Lett.} {\bf 81}, 3108-3111 (1998).

\bibitem{Hofstetter2002} Hofstetter, W., Cirac, J.~I., Zoller, P.,
  Demler, E. \& Lukin, M.~D. High-temperature superfluidity of fermionic
  atoms in optical lattices. \emph{Phys. Rev. Lett.} {\bf 89}, 220407
  (2002).

\bibitem{Greiner2002} Greiner, M., Mandel, O., Esslinger, T., H\"ansch,
  T.~W. \& Bloch, I. Quantum phase transition from a superfluid to a
  Mott insulator in a gas of ultracold atoms. \emph{Nature} {\bf 415},
  39-44 (2002).

\bibitem{Trebst2006} Trebst, S., Schollw\"ock, U., Troyer, M. \& Zoller,
  P. d-wave resonating valence bond states of fermionic atoms in optical
  lattices. \emph{Phys. Rev. Lett.} \textbf{96}, 250402 (2006).

\bibitem{Bloch2007} Bloch, I., Dalibard, J. \& Zwerger, W. Many-body
  physics with ultracold gases, \emph{Rev. Mod. Phys.} \textbf{80},
  885-964 (2008).

\bibitem{Stringari2008} Giorgini, L., Pitaevskii, L.~P. \& Stringari, S.
  Theory of ultracold Fermi gases.  \emph{Rev. Mod. Phys.} \textbf{80},
  1215-1275 (2008).
    
\bibitem{Georges2006} Georges, A. in \emph{Ultracold Fermi Gases} (eds
  Inguscio, M., Ketterle, W. \& Salomon, C.) 477-533 (IOS Press,
  Amsterdam, 2007).

\bibitem{Koehl2005} K\"ohl, M., Moritz, H., St\"oferle, T., G\"unter, K.
  \& Esslinger, T. Fermionic atoms in a three dimensional optical
  lattice: observing Fermi surfaces, dynamics, and interactions.
  \emph{Phys. Rev. Lett.} {\bf 94}, 080403 (2005).

\bibitem{Stoeferle2006} St\"oferle, T., Moritz, H., G\"unter, K.,
  K\"ohl, M. \& Esslinger, T. Molecules of fermionic atoms in an optical
  lattice, \emph{Phys. Rev. Lett.} {\bf 96}, 030401 (2006).

\bibitem{Chin2006} Chin, J.~K. et al.  Evidence for superfluidity of
  ultracold fermions in an optical lattice. \emph{Nature}, {\bf 443},
  961-964 (2006).

\bibitem{Rom2006} Rom, T. et al.  Free fermion antibunching in a
  degenerate atomic Fermi gas released from an optical lattice.
  \emph{Nature}, {\bf 444}, 733-736 (2006).

\bibitem{Strohmaier2007} Strohmaier, N. et al.  Interaction-controlled
  transport of an ultracold Fermi gas.  \emph{Phys. Rev. Lett.} {\bf
  99}, 220601 (2007).

\bibitem{Helmes2008} Helmes, R.~W., Costi, T.~A. \& Rosch, A. Mott
  transition of fermionic atoms in a three-dimensional optical trap,
  \emph{Phys. Rev. Lett.} {\bf 100}, 056403 (2008).

\bibitem{Rigol2003} Rigol, M., Muramatsu, A., Batrouni, G.~G. \&
  Scalettar, R.~T. Local quantum criticality in confined fermions on
  optical lattices. \emph{Phys. Rev. Lett.} {\bf 91}, 130403 (2003).

\bibitem{Georges1996} Georges, A., Kotliar, G., Krauth, W. \& Rozenberg,
  M.~J. Dynamical mean-field theory of strongly correlated fermion
  systems and the limit of infinite dimensions. \emph{Phys. Mod. Phys.}
  \textbf{68}, 13-125 (1996).

\bibitem{Koehl2006} K\"ohl, M. Thermometry of fermionic atoms in an
  optical lattice, \emph{Phys. Rev.~A} \textbf{73}, 031601(R) (2006).

\bibitem{Gebhard1997} Gebhard, F. \emph{The Mott metal-insulator
  transition -- models and methods}, (Springer, New York, 1997).

\bibitem{Gerbier2007} Gerbier, F. Boson Mott insulators at finite
  temperatures, \emph{Phys. Rev. Lett.} {\bf 99}, 120405 (2007).

\bibitem{Gerbier2006} Gerbier, F., F\"olling, S., Widera, A., Mandel, O.
  \& Bloch, I. Probing number squeezing of ultracold atoms across the
  superfluid-Mott insulator transition. \emph{Phys. Rev. Lett.}
  \textbf{96}, 090401 (2006).

\bibitem{Brinkman1970} Brinkman, W.~F. \& Rice, T.~M. Single-particle
  excitations in magnetic insulators, \emph{Phys. Rev. B.} \textbf{2},
  1324-1338 (1970).

\bibitem{Stoeferle2004} St\"oferle, T., Moritz, H., Schori, C., K\"ohl,
  M. \& Esslinger, T. Transition from a strongly interacting 1D
  superfluid to a Mott insulator. \emph{Phys. Rev. Lett.} {\bf 92},
  130403 (2004).
    
\bibitem{Kollath2007} Kollath, C., Iucci, A., McCulloch, I.~P. \&
  Giamarchi, T. Modulation spectroscopy with ultracold fermions in an
  optical lattice, \emph{Phys. Rev.~A} \textbf{74}, 041604(R) (2006).

\bibitem{Huber2008} Huber, S.~D., Theiler, B., Altman, E. \& Blatter, G.
  Amplitude mode in the quantum phase model, \emph{Phys. Rev.  Lett.}
  \textbf{100}, 050404 (2008).
   
\bibitem{Regal2003} Regal, C.~A. \& Jin, D.~S. Measurement of positive
  and negative scattering lengths in a Fermi gas of atoms.  \emph{Phys.
  Rev. Lett.} {\bf 90}, 230404 (2003).

\end{thebibliography}
\end{document}